\documentclass[]{emulateapj}

\shorttitle{Origin of 12\,$\mu$\MakeLowercase{m} Emission Across Galaxy
Populations from WISE and SDSS Surveys}
\shortauthors{Donoso et al.}

\begin{document}
\title{Origin of 12\,$\mu$\MakeLowercase{m} Emission Across Galaxy Populations
from WISE and SDSS Surveys}
\author{E. Donoso$^1$, Lin Yan$^1$, C. Tsai$^1$, P. Eisenhardt$^2$, D.
Stern$^2$, R. J. Assef$^{2,6}$, D. Leisawitz$^3$, T. H. Jarrett$^5$, S. A.
Stanford$^4$}
\affil{$^1$Spitzer Science Center, California Institute of Technology, 1200 E.
California Blvd., Pasadena CA 91125, USA}
\affil{$^2$Jet Propulsion Laboratory, California Institute of Technology,
Pasadena, CA 91109, USA}
\affil{$^3$Goddard Space Flight Center, Greenbelt, MD 20771, USA}
\affil{$^4$Department of Physics, University of California, Davis, CA 95616, USA}
\affil{$^5$Infrared Processing and Analysis Center, California Institute of
Technology, Pasadena, CA 91125, USA}
\affil{$^6$NASA Postdoctoral Program Fellow}

\begin{abstract}
We cross-matched Wide-field Infrared Survey Explorer (WISE) sources brighter
than 1 mJy at 12\,$\mu$m with the Sloan Digital Sky Survey (SDSS) galaxy
spectroscopic catalog to produce a sample of $\sim10^5$ galaxies at $\langle z
\rangle$\,=\,0.08, the largest of its kind. This sample is dominated
(70\%) by star-forming (SF) galaxies from the blue sequence, with total IR
luminosities in the range $\sim10^{8}-10^{12}\,L_\odot$.
We identify which stellar populations are responsible for most of the 12$\mu$m
emission. We find that most ($\sim$80\%) of the 12\,$\mu$m emission in SF
galaxies is produced by stellar populations younger than 0.6~Gyr. In contrast,
the 12\,$\mu$m emission in weak AGN ($L_{\rm[OIII]}<10^7\, L_\odot$) is produced
by older stars, with ages of $\sim 1-3$~Gyr. We find that $L_{\rm 12\mu m}$
linearly correlates with stellar mass for SF galaxies. At fixed 12\,$\mu$m
luminosity, weak AGN deviate toward higher masses since they tend to be hosted
by massive, early-type galaxies with older stellar populations. Star-forming
galaxies and weak AGN follow different $L_{\rm 12\mu m}$--SFR (star formation
rate) relations, with weak AGN showing excess 12\,$\mu$m emission at low SFR
($0.02-1\,M_\odot$\,yr$^{-1}$). This is likely due to dust grains heated by
older stars. While the specific star formation rate (SSFR) of SF galaxies is
nearly constant, the SSFR of weak AGN decreases by $\sim$3 orders of magnitude,
reflecting the very different star formation efficiencies between SF galaxies
and massive, early-type galaxies.
Stronger type II AGN in our sample ($L_{\rm[OIII]}>10^7\,L_\odot$), act as an
extension of massive SF galaxies, connecting the SF and weak AGN sequences.
This suggests a picture where galaxies form stars normally until an AGN
(possibly after a starburst episode) starts to gradually quench the SF activity.
We also find that 4.6--12\,$\mu$m color is a useful first-order indicator of SF
activity in a galaxy when no other data are available.
\end{abstract}


\section{Introduction}
A detailed picture of the present day galaxy populations, their evolution and
emission properties across different wavelengths is still far from complete.
Surveys like the Sloan Digital Sky Survey (SDSS; \citealt{york}) have collected
large amounts of information in the optical regime, while 2MASS
(\citealt{skrutskie}) and the IRAS mission (\citealt{neugebauer}) have
provided valuable, albeit relatively shallow, data sets from $J$
band up to 100~$\mu$m. More recently, the Wide-field Infrared Survey Explorer
(WISE; \citealt{wright}) has completed mapping the whole sky in the mid and far
infrared, at sensitivities much deeper than any other large-scale infrared
survey. For example, WISE is about 100 times more sensitive at 12\,$\mu$m than
IRAS.

While our understanding of the optical and far-IR properties of galaxies
(long-ward of 24$\mu$m) has grown steadily, thanks mainly to Spitzer and
Herschel, the spectral region between 10-15~$\mu$m remains comparatively
unexplored. In normal galaxies, the light at the redder optical bands and in the
$J$, $H$ and $K$ near-IR bands is closely tied to the total mass of the galaxy,
as it is dominated by the red population of older stars. At wavelengths longer
than $\sim$8~$\mu$m, emission from dust heated by younger stars becomes
increasingly relevant and begins to trace the star formation rate (SFR).

The rate at which galaxies transform gas into stars is one of the most
fundamental diagnostics that describes the evolution of galaxies. Of major
importance is to find what physical parameter(s) drive changes in the SFR. As
dust-reprocessed light from young stars is re-emitted mainly in the far-infrared
(FIR) regime, the FIR luminosity is one of the best tracers of star formation
activity (\citealt{kennicutt98}). It is well known that commonly used SFR
indicators, such as the UV continuum and nebular emission line fluxes, require
sometimes substantial corrections for dust extinction. Furthermore, these
corrections are highly uncertain and difficult to derive. For this reason,
integrated estimators based on the total infrared (IR) luminosity, either alone
or in combination with the ultraviolet luminosity (e.g. \citealt{heckman}), and
monochromatic estimators based mainly on 24~$\mu$m fluxes, alone or in
combination with H$\alpha$ luminosity (e.g. \citealt{wu};
\citealt{alonsoh}; \citealt{calzetti07}; \citealt{zhu}; \citealt{rieke};
\citealt{kennicutt09}), are increasingly being considered as reliable star
formation indicators for normal and dusty star-forming galaxies. The use of
any IR flux as a SFR indicator relies on the assumption that the IR continuum
emission is due to warm dust grains heated by young stars. However there is also
a contribution to dust reprocessed emission by old stellar populations,
associated more with the interstellar medium around evolved stars than to
recently born stars. In addition, some fraction of the IR luminosity may be
attributed to active galactic nuclei (AGN), if present (in
Section~\ref{sec:agn_effect} we find AGN emission is of minor importance in
most of our sample). The exact contribution of each component is difficult to
estimate without detailed spectral analysis.

 \citet{charyelbaz} and \citet{rieke} found correlations between the 12\,$\mu$m
luminosity and the total IR luminosity for small samples of nearby galaxies.
\citet{duc} found that sources in Abell 1689 with high ISO mid-IR color index
[15~$\mu$m]/[6.75~$\mu$m] are mostly blue, actively star forming galaxies, while
low mid-IR flux ratios correspond to passive early-type systems. They suggest
that 15~$\mu$m emission is a reliable indicator of obscured star
formation. Similarly, shorter wavelength mid-IR emission such as WISE
12\,$\mu$m is expected to be a practical tracer of star formation activity.

An important caveat is that far-infrared and/or radio measurements are only
available for a small fraction of known galaxies. Early work by
\citet{spinoglio} using IRAS all-sky data used 12\,$\mu$m to select unbiased
samples of active galaxies with fluxes above 220~mJy and to study their
luminosity function. Deep pencil-beam surveys have complemented these samples
with high redshift data. \citet{seymour} conducted a 12\,$\mu$m survey of the
ESO-Sculptor field (700~arcmin$^2$) with the ISO satellite down to 0.24 mJy.
\citet{roccav} used it to model mid-IR galaxy counts, revealing a population of
dusty, massive ellipticals in ultra luminous infrared galaxies (ULIRGS).
\citet{teplitz} performed imaging of the GOODS fields (150~arcmin$^2$) at
16~$\mu$m with the Spitzer spectrometer, finding that $\sim$15\% of objects are
potentially AGN at their depth of 40-65~$\mu$Jy. These surveys illustrate the
necessary tradeoff between depth and area covered, potentially limiting the
statistical significance of results due to cosmic variance. WISE provides the
data to significantly improve the situation. Our sample of $\sim$10$^5$ galaxies
(see below) is over 200 times more sensitive than IRAS-based surveys and covers
an area $\sim$370 times larger than the GOODS samples.

In this work we explore the physical properties of 12\,$\mu$m-selected galaxies
in the local Universe, using a large sample of star forming galaxies and AGN
with available redshifts and emission line measurements from the SDSS. This is
by far
the largest 12\,$\mu$m sample to date, and we use it to trace the origin of
IR emission across different galaxy populations and to investigate how IR
emission relates to stellar mass. We also explore using 12\,$\mu$m
luminosity as a proxy for SFR to distinguish intense starburst activity from
quiescent star formation. Since we employ the SDSS spectroscopic catalog, our
results apply to relatively bright galaxies at low redshift. WISE certainly
recovers other populations of galaxies, ranging from low metallicity blue
compact dwarf galaxies at very low redshift (\citealt{griffith}) to highly
obscured sources at high redshift (\citealt{eisenhardt}). \citet{lake} shows
that WISE detects $L^*$ galaxies out to $z\sim0.8$ in the 3.4$\mu$m band, while
\citet{stern2011} shows that WISE is a very capable AGN finder, sensitive to
both obscured and unobscured QSOs. A companion paper by \citet{yan} analyzes
more diverse galaxy populations observed by WISE and SDSS (including deeper
photometric SDSS objects), while here we focus on 12\,$\mu$m-selected sources
with available spectra.

This paper is organized as follows. In Section 2 we describe the surveys used in
this work as well as explain the construction of our joint WISE-SDSS sample. In
Section 3 we characterize the galaxy populations and present the results on the
analysis of the mid-IR emission and SFR. Finally, Section 4 summarizes our
results and discusses the implications of this work.

Throughout the paper we assume a flat $\Lambda$CDM cosmology, with
$\Omega_{m}=0.3$ and $\Omega_{\Lambda}=0.7$. We adopt
$H_{0}$=70~km~s$^{-1}$~Mpc$^{-1}$.

\begin{figure*}
\epsscale{1.0}
\plotone{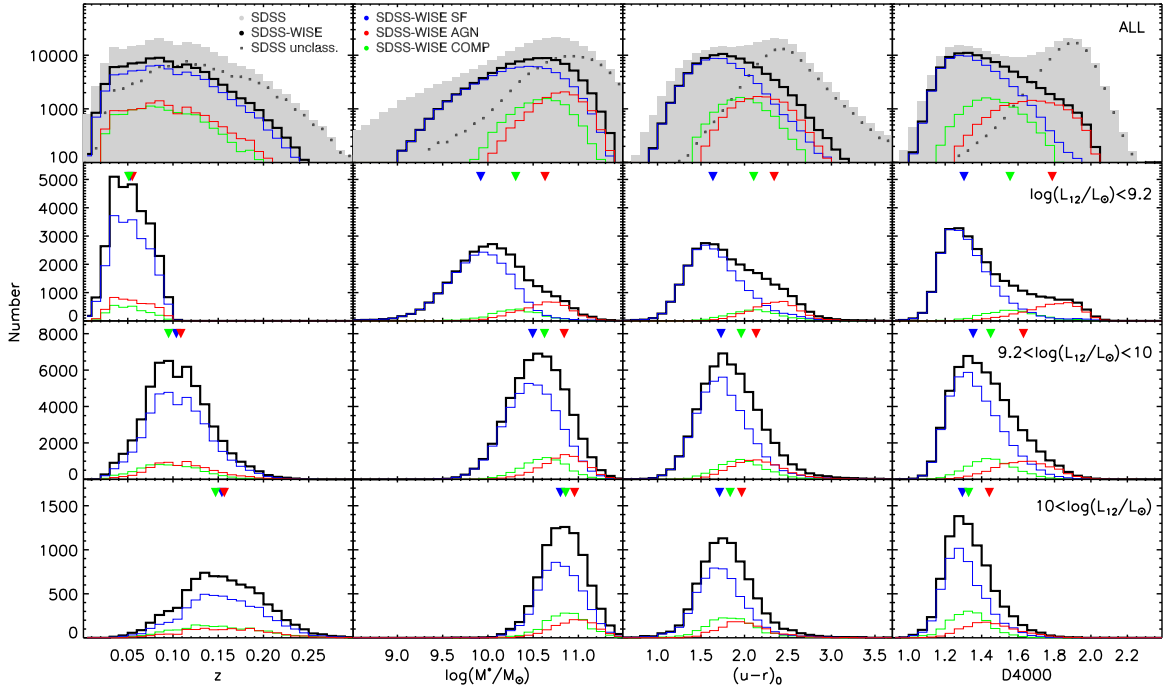}
\caption{Distributions of redshift, stellar mass,restframe color $(u-r)_0$ and
D$_{4000}$ for all WISE-SDSS 12\,$\mu$m-selected galaxies (black), star forming
galaxies (blue), AGN (red) and composite galaxies (green). We also show the
distribution for all optical MPA-JHU galaxies (gray), as well as for galaxies
without BPT classification due to the lack of detected emission lines (dotted).
The top row shows the complete sample while the lower three rows split into low,
intermediate and high IR luminosity. Triangle markers indicate the means of the
respective distributions.}
\label{fig:dist_lumw3}
\end{figure*}

\section{Data}
\subsection{The Wide-field Infrared Survey Explorer Catalog}
WISE has mapped the full sky in four bands centered at 3.4, 4.6, 12 and
22~$\mu$m, achieving 5$\sigma$ point source sensitivities better than 0.08,
0.11, 1 and 6~mJy, respectively. Every part of the sky has been observed
typically around 10 times, except near the ecliptic poles where the coverage is
much higher. Astrometric precision is better than 0.15$^{\prime\prime}$ for high
S/N sources (\citealt{jarrett}) and the angular resolution is 6.1, 6.4, 6.5 and
12$^{\prime\prime}$ for bands ranging from 3~$\mu$m to 22~$\mu$m. 

This paper is based on data from the WISE Preliminary Release 1 (April 2011),
which comprises an image atlas and a catalog of over 257 million sources from
57\% of the sky. An object is included in this catalog if it:
(1) is detected with SNR$\geq$7 in at least one of the four bands, (2) is
detected on at least five independent single-exposure images in at least one
band, (3) has SNR$\geq$3 in at least 40\% of its single-exposure images in one
or more bands, (4) is not flagged as a spurious artifact in at least one band.
We refer the reader to the WISE Preliminary Release Explanatory Supplement for
further details\footnote{WISE data products and documentation are available at
\\
http://irsa.ipac.caltech.edu/Missions/wise.html} (\citealt{cutri}).

\subsection{The MPA-JHU Sloan Digital Sky Survey Catalog}
The Sloan Digital Sky Survey (\citealt{york}; \citealt{stoughton}) is a
five-band photometric ($ugriz$ bands) and spectroscopic survey that has mapped
a quarter of the sky, providing photometry, spectra and redshifts for
about a million galaxies and quasars. The MPA-JHU catalog (\citealt{brinchman},
hereafter B04) is a value-added catalog based on data from the Seventh Data
Release (DR7, \citealt{abazajian}) of the SDSS\footnote{The MPA-JHU catalog is
publicly available at\\ http://www.mpa-garching.mpg.de/SDSS/DR7/}. It consists
of almost $10^6$ galaxies with spectra reprocessed by the MPA-JHU team, for
which physical properties based on detailed emission line analysis are readily
available. Here we give a brief description of the catalog and the methodology
employed to estimate SFRs. We refer the reader to the original papers for an
in-depth discussion.

The MPA-JHU catalog classifies galaxies according to their emission lines,
given the position they occupy in the BPT (\citealt{bpt}) diagram that plots
the [OIII]~$\lambda$5007\AA/H$\beta$ versus [NII]~$\lambda$6584\AA/H$\alpha$
flux ratios. This separates galaxies with different ionizing sources as
they populate separate sequences on the BPT diagram. In most galaxies, normal
star formation can account for the flux ratios. However, in some cases an
extra source such as an AGN is required. In this paper we follow this BPT
classification to distinguish between: \textbf{(i) star-forming galaxies} (class
SF and LOW S/N SF from B04), \textbf{(ii) active galactic nuclei} (class AGN and
Low S/N LINER from B04), and \textbf{(iii) composite systems} that present
signatures of the two previous types (class C from B04). Note that broad-lined 
AGN like quasars and Seyfert 1 galaxies are are not included in the sample, as
they were targeted by different criteria by the SDSS.

Star formation rates are derived by different prescriptions depending on the
galaxy type. The methodology adopted by B04 is based on modeling emission
lines using \citet{bruzual93} models along with the CLOUDY photoionization model
(\citealt{ferland}) and spectral evolution models from \citet{charlot08}
to subtract the stellar continuum. To correct lines for dust attenuation,
MPA-JHU adopts the multicomponent dust model of \citet{charlot_fall}, where
discrete dust clouds are assumed to have finite lifetime and a realistic spatial
distribution. This approach produces SFRs that take full advantage of all
modeled emission lines. For AGN and composite galaxies in the sample, SFRs were
estimated by the relationship between the D$_{4000}$ spectral index and the
specific SFR (SFR/M$_{\odot}$ or SSFR), as calibrated for star forming galaxies
(see Fig. 11 in B04). These estimates have been corrected in the latest MPA-JHU
DR7 release by using improved dust attenuations and improved aperture
corrections for non-SF galaxies following the work by \citet{salim07}. Gas-phase
oxygen abundances, 12+log(O/H), are available for star forming galaxies as
calculated by \citet{tremonti}. Thoughout this paper we adopt the spectral line
measurements as well as estimates of SFR, metallicity and dust extinction given
by the MPA-JHU catalog.

The SDSS pipeline calculates several kinds of magnitudes. In this work we have
adopted modified Petrosian magnitudes for flux measurements, which capture a
constant fraction of the total light independent of position and distance. When
appropriate, we have also used model magnitudes (\textit{modelMag}) as they
provide the most reliable estimates of galaxy colors. Magnitudes are corrected
for galactic reddening using the dust maps of \citet{schlegel}.

\begin{figure*}
\epsscale{0.9}
\plotone{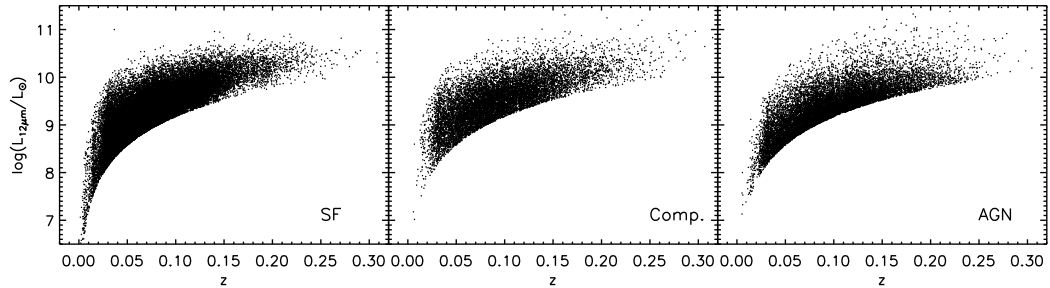}
\caption{Distribution of 12\,$\mu$m luminosity for galaxies separated according
to spectral type as: SF galaxies (left), composite systems (middle),
and AGN (right).}
\label{fig:zlumw3}
\end{figure*}

\subsection{The Joint WISE-SDSS Galaxy Sample}
We have crossmatched data from WISE and the MPA-JHU catalog to construct a
galaxy sample covering an effective area of 2344 deg$^2$, or 29\% of the DR7
(legacy) spectroscopic footprint. WISE sources were selected to have 12\,$\mu$m
fluxes above 1~mJy, also requiring objects to have clean photometry at 3.4, 4.6
and 12\,$\mu$m. For MPA-JHU sources, we selected
objects with de-reddened Petrosian magnitude $r_{\rm petro}<17.7$ and
$r$-band surface brightness $\mu_{r}<23$~mag~arcsec$^{-2}$. This selects a
conservative version of the SDSS main galaxy sample (see \citealt{strauss}).
Using a matching radius of 3$^{\prime\prime}$ we find 96,217 WISE objects with
single optical matches (40\% of the SDSS sample in the intersection area), and
73 sources with two or more counterparts. The latter are mostly large extended
sources or close interacting systems of two members. For the rest of this paper
we will focus on the single IR-optical matches that constitute the vast majority
($>$99.9\%) of the galaxy population. By using random catalogs generated over
the effective area, the expected false detection fraction at 3$^{\prime\prime}$
is 0.05\%.

Each region of the sky has been observed at least 10 times by WISE, with the
number of observations increasing substantially toward the ecliptic poles.
Within our effective area, the median coverage depth at 12\,$\mu$m is about 13,
varying tipically between 10 and 20. At these levels, the average completeness
of the sample is expected to be over 90\%, as shown in the WISE Preliminary
Release Explanatory Supplement (Sec. 6.6).

\section{Analysis and Results}
\subsection{Derived Properties}
We derived stellar masses for all galaxies using the \textit{kcorrect} algorithm
(\citealt{blanton}), which fits a linear combination of spectral templates to
the flux measurements for each galaxy. These templates are based on a set of
\citet{bruzual03} models that span a wide range of star formation histories,
metallicities and dust extinction. This algorithm yields stellar masses that
differ by less than 0.1 dex on average from estimates using other methods (for
example, the method based on fitting the 4000{\AA} break strength and $H\delta$
absorption index proposed by \citealt{kauff03}). 

To derive restframe colors and monochromatic luminosities in the infrared we
used the fitting code and templates\footnote{Templates and code are available
at\\http://www.astronomy.ohio-state.edu/$\sim$rjassef/lrt/} of \citet{assef},
applied to our combined \emph{ugriz} photometry plus WISE 3.4~$\mu$m,
4.6~$\mu$m and 12\,$\mu$m fluxes. \citet{assef} present a set of low-resolution
empirical spectral energy distribution (SED) templates for galaxies and AGN
spanning the wavelength range from 0.03~$\mu$m to 30~$\mu$m, constructed with
data from the NOAO Deep Wide-Field Survey Bo\"{o}tes field (NDFWS, 
\citealt{jannuzi}) and the AGN and Galaxy Evolution Survey (AGES, 
\citealt{kochanek}). The code fits three galaxy SED templates that represent
an old stellar population (elliptical), a continuously star-forming population
(spiral) and a starburst component (irregular), plus an AGN SED template with
variable reddening and IGM absorption. These templates have been successfully
used to test the reliability of IRAC AGN selection, and to predict the
color-color distribution of WISE sources (\citet{assef}). We also use these
templates to assess the relative contribution of AGN to the energy budget.

Instead of trying to derive a new calibration of the SFR in the IR, we take the
approach of comparing IR luminosities directly to optical dust-corrected SFRs.
This makes our results largely independent of any particular SFR calibration.
Note all optical SFRs used in this paper have been corrected for dust
extinction.

\begin{figure}
\epsscale{1.0}
\plotone{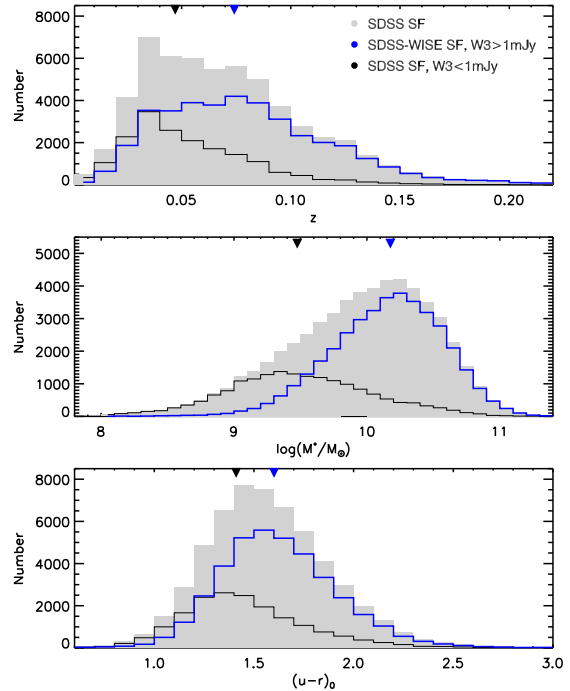}
\caption{Distribution of redshift, stellar mass and restframe color $(u-r)_0$
for WISE-SDSS 12\,$\mu$m-selected galaxies classified as star forming (blue). We
also show the distributions for all optical star-forming galaxies (gray) and for
star-forming galaxies without 12\,$\mu$m flux densities above 1~mJy (black).
Triangle markers indicate the means of the distributions.}
\label{fig:dist_sf}
\end{figure}

\begin{figure*}
\epsscale{1.0}
\plotone{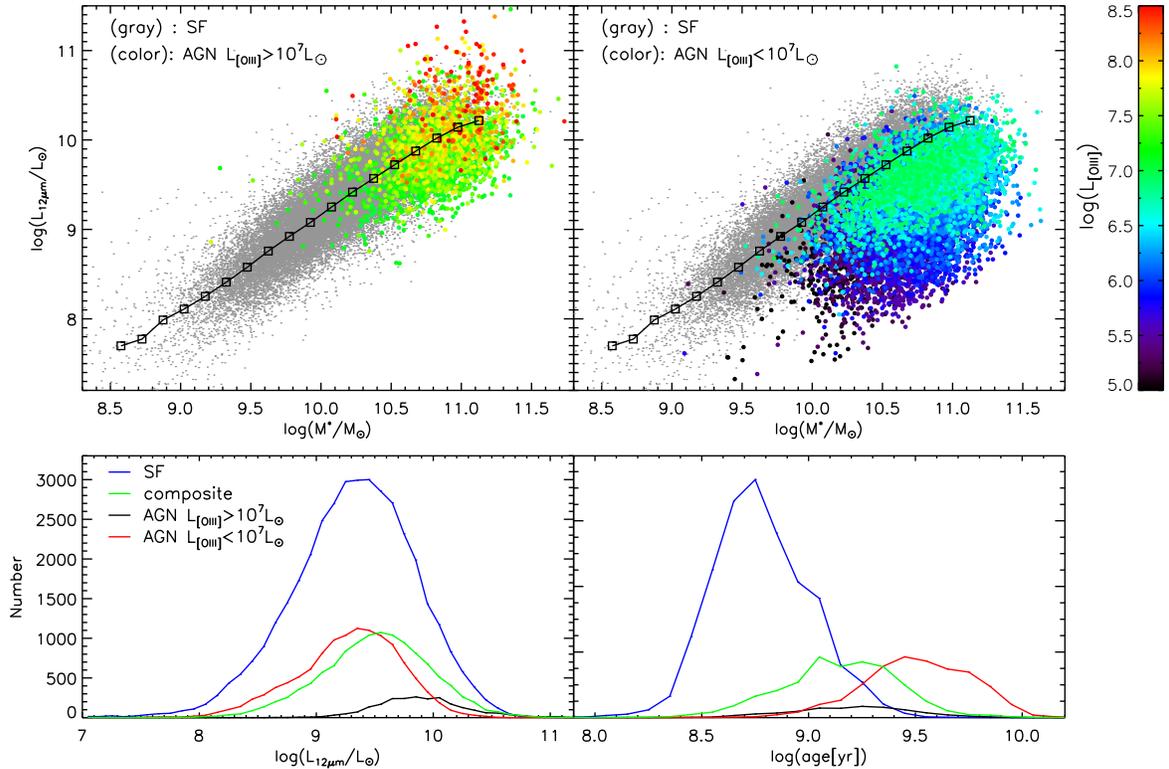}
\caption{12\,$\mu$m infrared luminosity as a function of stellar mass for SF
galaxies (gray), strong AGN with $L_{\rm [OIII]}>10^{7}L_{\odot}$ (color points,
top left), and weak AGN with $L_{\rm [OIII]}<10^{7}L_{\odot}$ (color points, top
right). The solid line shows the median $L_{\rm 12\mu m}$ of SF galaxies as a
function of mass. Bottom panels show the distribution of $L_{\rm 12\mu m}$ and
stellar age for the various populations as indicated.}
\label{fig:mass_lumw3_sptype}
\end{figure*}

\subsection{General Properties of 12\,$\mu$\MakeLowercase{m} Galaxies}
We begin our analysis by exploring the general properties of the WISE 
12\,$\mu$m-selected galaxy sample. The sample is composed of a mixture
of 70\% SF galaxies, 15\% AGN, 12\% composite galaxies and 3\% galaxies lacking
BPT classification due to the absence of detectable lines in the spectra
(most lack H$\alpha$ emission). The composition of the MPA-JHU optical sample
is 44\%~SF, 12\% AGN, 6\% composite, 37\% unclassifiable, which means that the
12\,$\mu$m selection is highly efficient in recovering SF systems and avoids
objects with weak or no emission lines. In terms of the total optical galaxy
populations, 61\% (SF), 53\% (AGN) and 76\% (composite) of the SDSS galaxies
have 12\,$\mu$m flux densities above 1~mJy. In Figure~\ref{fig:dist_lumw3}
(top row) we plot the distribution of redshift, stellar mass, D$_{4000}$ index
and restframe $u-r$ color for SF galaxies, AGN and composite systems, as
well as for the three classes all together. The majority of WISE-SDSS 12\,$\mu$m
sources are SF galaxies at $\langle z\rangle=0.08$ with stellar masses of $\sim
10^{10.2}$M$_{\odot}$; these are typical values for the SF class. They clearly
populate the blue peak of the galaxy bimodal distribution around $(u-r)_{0}=1.6$
and have inherently young stellar populations (D$_{4000}\sim$1.3, or ages of
$\sim$0.5~Gyr). AGN are, as expected, comparatively more massive
(M$^{*}\sim10^{10.7}$M$_{\odot}$), redder ($(u-r)_{0}\sim 2.1$) and older
($\sim$1-6~Gyr), dominating the massive end of the 12\,$\mu$m galaxy
distribution. As a population, AGN do not differ significantly (in terms of
these properties) in comparison to the corresponding purely optical sample.
Composite galaxies present intermediate properties between the SF and AGN
samples. Note that the bulk of galaxies lacking BPT classification
(due to weak or absent emission lines) is missed in our IR-optical sample.
These objects primarily populate the red sequence of the galaxy distribution
(e.g. \citealt{baldry}) and hence are not expected to be prominent at
12\,$\mu$m.

We then divide the sample into three subsamples of monochromatic infrared
luminosity: faint ($L_{\rm 12\mu m}<10^{9.2}L_{\odot}$), intermediate
($10^{9.2}L_{\odot}<L_{\rm 12\mu m}<10^{10}L_{\odot}$), and bright
($L_{\rm 12\mu m}>10^{10}L_{\odot}$) sources. There are no large biases with
redshift, i.e. the different classes are sampled roughly without preference at
all luminosities. Both SF galaxies and AGN become more massive for higher IR
luminosities and we have checked that this also holds in narrow redshift slices.
As we will show below, this is due to the coupling between the IR and the
optical emission. Interestingly, the 12\,$\mu$m SF galaxies change by a factor
of 0.8~dex in mass
toward high 12\,$\mu$m luminosities while keeping the same color and stellar
content. The AGN population, while getting slightly more massive, becomes bluer
and dominated by younger stars as IR luminosity increases.
Figure~\ref{fig:zlumw3} shows the redshift distribution of the restframe
12\,$\mu$m luminosity for our sample. It can be seen that while high luminosity
sources naturally lie at higher redshifts, the redshift distribution of the
different classes is very similar, except at the lowest redshifts ($z<0.02$)
where very few AGN/composite galaxies are observed.

The absence of red sequence galaxies in our sample is not surprising, but
there is also a number of SF galaxies without IR emission due to our flux limit.
Figure~\ref{fig:dist_sf} shows the mass, redshift and colors of SF galaxies with
and without 12\,$\mu$m emission, as well as for the entire SF optical sample. On
average, SF galaxies not present in our sample have bluer colors, lie at lower
redshifts, and have stellar masses around $10^{9.3}$M$_{\odot}$, roughly an
order of magnitude below the 12\,$\mu$m SF galaxy sample. At this level,
galaxies have very little dust mass, and hence can not re-radiate much in the
IR.

The main result here is that WISE 12\,$\mu$m-selected galaxies are primarly
typical blue sequence (SF) galaxies. It is safe to assume that the majority of
blue sequence galaxies correspond to late morphological types (e.g.
\citealt{strateva}; \citealt{shimasaku}; \citealt{baldry}). AGN and composite
objects are also represented, belonging either to the red sequence or to a
transitional regime
among the two former classes. It is interesting that the detection efficiency is
largest for composite systems, which was also found by \citet{salim09} for
24~$\mu$m sources that lie in the so called \textit{green valley} (e.g.
\citealt{martin}). This is a region located between the red and blue cloud
sequences, best identified in the NUV-r color-magnitude diagram, where
SF activity is being actively quenched and galaxies are thought to be in
transitional stage in their migration from the blue to the red sequence.

\begin{figure}
\epsscale{1.0}
\plotone{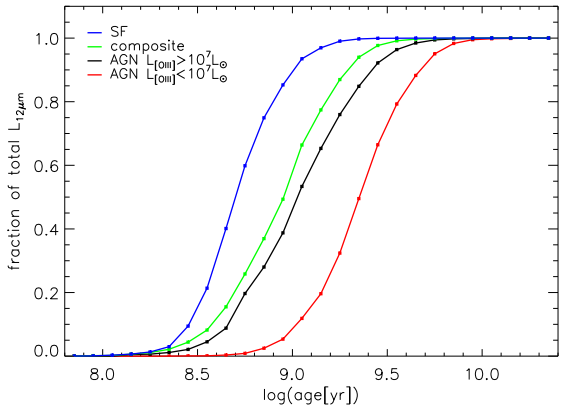}
\caption{Cumulative fraction of integrated 12\,$\mu$m luminosity for galaxies
dominated by stellar populations of a given age, separated according to
spectral type as: SF galaxies (blue), composite systems (green), strong AGN
(black) weak AGN (red).}
\label{fig:lumage}
\end{figure}

\begin{figure*}
\epsscale{1.1}
\plotone{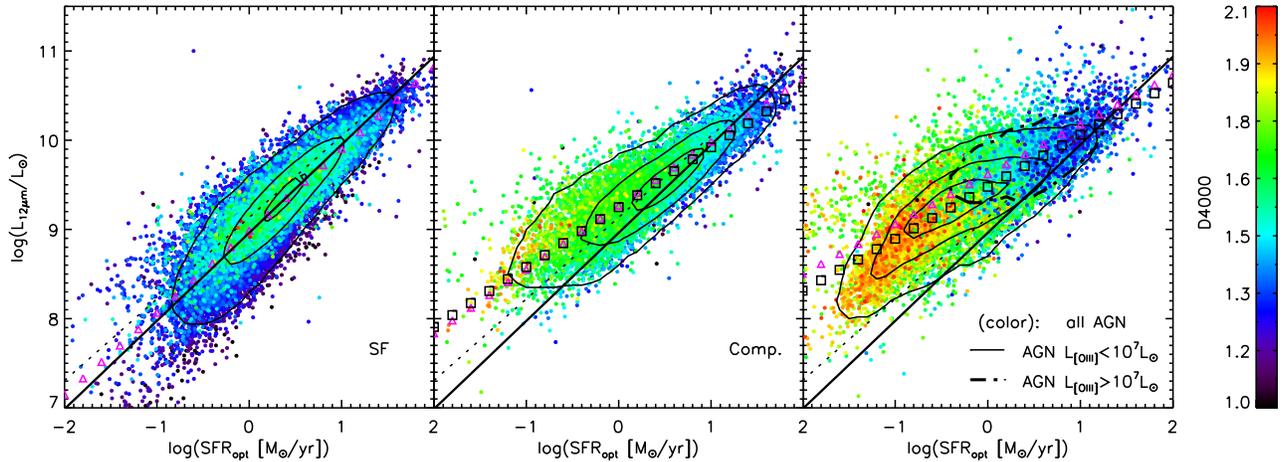}
\caption{Infrared luminosity at 12\,$\mu$m as a function of dust-corrected star
formation rate derived from optical emission lines, for galaxies classified as
SF (left), Composite (middle) and strong/weak AGN (right). Linear fits are
shown for the cases of: SF galaxies (solid line, all panels), composite systems
(squares, middle panel) and AGN (squares, right panel). The dashed line is the
\citet{charyelbaz} conversion between SFR and 12$\mu$m luminosity. For sources
brighter than 5~mJy at $22\mu$m, we also plot the corresponding $L_{\rm 22\mu
m}$--SFR relationships (triangles). In all panels, the mean stellar age of the
dominant stellar population is indicated by the D$_{4000}$ color scale on the
right. Contours enclose 95\%, 68\% and 33\% of the density distribution (68\%
for the case of strong AGN).}
\label{fig:ltir_sfr_d4000}
\end{figure*}

Most galaxies in our sample are either normal luminosity IR galaxies
($L_{\rm 12\mu m}\sim10^{9.2-10}$\,L$_{\odot}$; 60\%), low luminosity IR
galaxies ($L_{\rm 12\mu m}<10^{9.2}$\,L$_{\odot}$; 31\%) or luminous infrared
galaxies (LIRGs; $L_{\rm 12\mu m}\sim10^{10-10.8}$\,L$_{\odot}$; 9\%). However,
ULIRGs are also present. There are 114 objects with $L_{\rm 12\mu
m}>10^{10.8}$\,L$_{\odot}$ (roughly equivalent to $L_{\rm
TIR}>10^{12}$\,L$_{\odot}$ using the conversion of
\citealt{charyelbaz}), corresponding to a surface density of 0.049~deg$^{-2}$.
This is comparable to the 0.041~deg$^{-2}$ surface density found by \citet{hou}.
These ULIRGs naturally lie at higher redshift ($\langle z \rangle=0.2$) and
populate the massive end of the SF sequence above $\sim 10^{11}$\,M$_{\odot}$.
We reiterate that these results come from matching WISE to the relatively
bright SDSS spectroscopic sample. WISE galaxy populations down to $r\sim$22.6
are analyzed in \citet{yan}.

\subsection{12\,$\mu$m Luminosity and Stellar Mass}
\label{sec:lumass}
We now have a large sample of 12\,$\mu$m-selected galaxies that range from
low-IR to ULIRG luminosities, for which high quality optical spectra and
dust-corrected optical SFRs are available. First we examine the relation between
$L_{\rm 12\mu m}$ and stellar mass. B04 have shown that, at least for star
forming systems, SFR and stellar mass are strongly correlated in the local
universe. There is also evidence that this relationship evolves with redshift
(\citealt{noeske}, \citealt{daddi}). Although it is expected that more massive
galaxies are naturally more luminous, it is unclear whether more massive systems
would have more dust emission in the mid-IR. Figure ~\ref{fig:mass_lumw3_sptype}
shows the correlation between monochromatic 12\,$\mu$m luminosity and stellar
mass for our sample. The correlation is tight for SF systems over nearly three
orders of magnitude in stellar mass (gray points, top panels). Several studies
have found that the distributions of [OIII] emission line flux to the AGN
continuum flux at X-ray, mid/far-IR and radio wavelengths (i.e. where
stellar emission and absorption by the torus are least significant) are very
similar for both type I and type II AGN (\citealt{mulchaey}; \citealt{keel}; 
\citealt{alonsoh97}). Based on this, \citet{kauff03agn} and \citet{heckman04}
have shown that [OIII] flux is a reliable estimator of AGN activity. Following
these works, we split the AGN sample by [OIII] luminosity. We see that strong
AGN ($L_{\rm [OIII]}>10^{7}$\,L$_{\odot}$) have IR luminosities considerably
larger than weak AGN ($L_{\rm [OIII]}<10^{7}$\,L$_{\odot}$), following
approximately the same relationship with mass as SF systems. Weak AGN, instead,
lie well below that correlation and show a larger scatter. We note that the
contribution by star forming regions to the [OIII] flux is $<$7\%
(\citealt{kauff03agn}).

In the right bottom panel of Figure~\ref{fig:mass_lumw3_sptype} we compare the
ages of stars in all subsamples as derived from the D$_{4000}$ spectral index.
SF galaxies have the youngest stellar populations, peaking at $\sim$0.5~Gyr,
followed by considerably older composite galaxies ($\sim$1.5-2~Gyr). Strong
AGN, which dominate the massive end, have intermediate stellar populations
($\sim$1.5~Gyr) that are closer to SF/Composite systems than to weak AGN (see
Fig. \ref{fig:dist_lumw3}). In contrast, weak AGN tend to be hosted by
early-type galaxies with significantly older stars ($\sim$3~Gyr) as found also
by \citet{kauff03agn} after comparing AGN host sizes, surface densities and
concentration ratios with those of normal early-type galaxies. This highlights
the importance that young/old stars have in powering the 12\,$\mu$m
emission. For AGN of roughly similar stellar mass, only when younger stars begin
to dominate (and the active nuclei becomes more powerful) is the IR emission
comparable to actively star-forming systems. Qualitatively, the same result
holds if we use the dust-corrected $r$-band absolute magnitude instead of
stellar mass.

Since galaxies of different ages have very different IR output, an interesting
question to address is how the 12\,$\mu$m luminosity budget depends upon the age
of the stellar populations. Figure~\ref{fig:lumage} shows the cumulative
fraction of the \emph{integrated} 12\,$\mu$m luminosity produced in
galaxies of a given age. In SF galaxies, $\sim$80\% of the total IR luminosity
is produced by galaxies younger than 0.6~Gyr. Composite galaxies and strong AGN
reach the same fraction at ages of 1.5~Gyr and 2~Gyr, respectively. In
weak AGN, instead, most of the mid-IR emission is produced within a range of
$\sim$1-3~Gyr. This inventory of 12\,$\mu$m luminosity in the local universe
shows where the bulk of the IR emission resides, shifting from stellar
populations of a few hundred Myr in actively SF galaxies to a few Gyr in
galaxies hosting weak AGN. Thus, it underlines again the important role that
young/old stars have in powering 12\,$\mu$m emission. As we will see later
in Section \ref{sec:ssfr}, this also supports the idea that transition galaxies
(i.e. composite/strong AGN) form a smooth sequence that joins highly active
galaxies with quiescent galaxies.

\subsection{12\,$\mu$m Luminosity and Star Formation Rate}
We now explore the relationship between infrared luminosity and optical,
dust-corrected SFR. Figure~\ref{fig:ltir_sfr_d4000} shows $L_{\rm 12\mu m}$
versus $SFR_{\rm opt}$, color-coded by the D$_{4000}$ spectral index. As
discussed by \citet{kauff03}, the D$_{4000}$ index is a good indicator of the
mean age of the stellar population in a galaxy. The dashed line indicates the
reference relation of \citet{kennicutt98}, as given by \citet{charyelbaz} in
terms of 12\,$\mu$m luminosity, to convert IR
luminosity into ``instantaneous'' SFR. It was derived from simple theoretical
models of stellar populations with ages 10-100~Myr without considering factors
like metallicity or more complex star formation histories. While this makes it
strictly valid only for young starbursts, the Kennicutt relation is quite often
applied to the more general population of star forming galaxies.
Figure~\ref{fig:ltir_sfr_d4000} shows that the IR emission from SF galaxies
(left panel) correlates fairly well with $SFR_{\rm opt}$. The correlation is
tighter for high SFRs becoming broader and slightly asymmetric for low SFRs. A
least-squares fit to the SF sample is given by
\begin{equation}
\log L^{\rm SF}_{\rm 12\mu m}=(0.987\pm0.002)\log SFR_{\rm opt}+(8.962\pm0.003).
\end{equation}
This expression is close to the \citet{charyelbaz} relation at high SFRs, which
is not surprising given that relation was calibrated using the IRAS Bright
Galaxy Sample (\citealt{soifer}), i.e. luminous galaxies with $L_{\rm 12\mu
m}>10^{9}$\,L$_{\odot}$. The small differences are likely attributable to
luminosity/redshift selection effects and the slight differences
between ISO and WISE $12\mu$m filters. More importantly, the agreement is quite
good considering the different origin (FIR vs optical emission lines) of the
SFRs. Relative to the \citet{charyelbaz} conversion, $L_{\rm 12\mu m}$ is
comparatively suppressed by a factor $>1.6$ for $SFR_{\rm opt}$ below
$\sim$0.1\,M$_{\odot}$\,yr$^{-1}$. This is likely because low SFR systems have
very low stellar masses ($<10^9$\,M$_{\odot}$), and therefore become more
transparent due to the increasing fraction of stellar light that escapes
unabsorbed by dust. We note, however, that the spatial distrbution of dust in
HII regions and/or molecular clouds could also have influence (e.g.
\citealt{leisawitz}). In addition, we have repeated the test for the bluest
galaxies in the SF galaxy class, obtaining no significant differences. This
suggests that other effects like metallicity could also be relevant.

For AGN, the coupling between optical SFR and IR luminosity follows a different
relationship (right panel of Figure~\ref{fig:ltir_sfr_d4000}). Most AGN lie in a
broader distribution \emph{above} the instantaneous conversion of Kennicutt,
particularly those with $SFR_{\rm opt}$ below $\sim$1\,M$_{\odot}$\,yr$^{-1}$.
For a fixed SFR, the IR emission is higher by a factor of several relative to SF
galaxies, suggesting that $L_{\rm 12\mu m}$ is not driven by the current SF for
most AGN. Weak AGN are predominantely associated with massive, early-type
galaxies increasingly dominated by old stars at low SFRs 
($\sim$0.1\,M$_{\odot}$\,yr$^{-1}$). Given their red optical colors, it is
unlikely that recent SF could be responsible for their IR emission. More likely,
dust grains heated due to older stars or an AGN are driving this emission. Note,
however, that in Section~\ref{sec:agn_effect} we will show that the contribution
of AGN at 12\,$\mu$m is of minor importance for most AGN, except perhaps for
most powerful ones. Only strong AGN, which are dominated by
intermediate-to-young stellar populations, tend to occupy a region similar to SF
galaxies in Figure~\ref{fig:ltir_sfr_d4000}. They show a clear ``excess'' in
$L_{\rm 12\mu m}$ at $SFR_{\rm opt}\sim$0.5\,M$_{\odot}$\,yr$^{-1}$ that
gradually decreases when stars get younger toward higher SFRs. This shows that
the age of stars in an AGN is an important factor in determining the origin of
the 12\,$\mu$m emission. An expression fitting AGN (weak and strong) is given by
\begin{equation}
\log L^{\rm AGN}_{\rm 12\mu m}=(0.582\pm0.004)\log SFR_{\rm
opt}+(9.477\pm0.002).
\end{equation}
Recent work by \citet{salim09} compared NUV/optical SFRs with $L_{\rm TIR}$
calibrated from 24~$\mu$m fluxes for red and green sequence objects at
$z\sim0.7$ (corresponding to restframe 14\,$\mu$m, close to the WISE 12\,$\mu$m
band). They find large excess IR emission for a given SFR, attributed mainly to
older stellar populations, and to a lesser extent to an AGN. We find broadly
consistent results, but for 12\,$\mu$m-selected AGN sources with optical SFRs.
\citet{kelson} have suggested that thermally pulsating AGB carbon stars (TP-AGB)
with ages of 0.2-1.5~Gyr (corresponding to D$_{4000}\sim$1.2-1.5) can also
contribute significantly to the mid-IR flux. As seen in Section
\ref{sec:lumass}, this does not seem to be important for our much older, weak
AGN, and perhaps marginally relevant in strong AGN that have typical stellar
ages slightly above the upper 1.5~Gyr limit. The case of SF galaxies is
interesting because most of the 12~$\mu$m luminosity seems to originate from
galaxies in the $\sim$0.3-0.6~Gyr age range and this luminosity correlates
relatively well with the optical SFR. While the age ranges seems to overlap, it
is difficult to prove whether TP-AGB dominate the emission or not. Further SED
decomposition and modeling of stellar populations is required to find the
fraction of mid-IR luminosity powered by TP-AGB stars, a task that is
potentially complicated by metallicity effects and uncertainties in the ensemble
colors of such stars. However, the general picture is consistent with previous
results (\citealt{salim09}; \citealt{kelson}) that find evidence for the mid-IR
being sensitive to star formation over relatively long ($>$1.5~Gyr) timescales.

Finally, we consider composite systems, which present considerable SF activity
along with spectral signatures of an AGN (middle panel of
Figure~\ref{fig:ltir_sfr_d4000}). By definition, these objects have up to
40\% (see B04) of their H$\alpha$ emission coming from a non-stellar origin,
though the fraction is $<$15\% for most galaxies. With masses, ages and optical
colors intermediate between the SF and AGN sequences, composite galaxies closely
follow the Kennicutt relation, except at the low SFR end where older stars once
again begin to dominate. A least-squares fit for composite galaxies has a slope
intermediate between AGN and SF galaxies, and is given by
\begin{equation}
\log L^{\rm comp}_{\rm 12\mu m}=(0.671\pm0.003)\log SFR_{\rm
opt}+(9.249\pm0.002).
\end{equation}
We note that the optical SFRs utilized here, while not ideal, are probably the
best estimates publicly available for such a large and diverse population of
galaxies in the local universe. Other methodologies that use more sophisticated
dust corrections and employ H$_2$, FIR or radio data could provide more
accurrate values, but are difficult to apply across the entire sample and data
are not always available (see \citealt{saintonge} for a calibration of SFR based
on H$_2$ masses). This is particularly relevant for SFRs in AGN, which can
present large $>$0.5 dex formal errors (see Figure 14 of \citealt{brinchman}).
We have verified that the SDSS SFRs in our sample are in broad agreement with
UV-based SFRs derived by \citet{salim07} (S. Salim, private communication), with
an average offset/scatter of 0.055/0.387 for the total sample (0.013/0.334 for
strong AGN, 0.242/0.567 for weak AGN).

\subsection{22~$\mu$m Luminosity and Star Formation Rate}
WISE is less sensitive at 22~$\mu$m than at 12\,$\mu$m, but a significant
fraction of our sample ($\sim$30\%) has measured 22~$\mu$m fluxes
above 5~mJy (note we find no 22~$\mu$m galaxies without 12\,$\mu$m
detection). Similar to the 12\,$\mu$m sample, this 22~$\mu$m subsample is a
mixture comprised of 65\%~SF galaxies, 14\%~AGN and 19\%~composite systems.
However, the fraction of strong AGN is 38\%, compared to 16\% for 12\,$\mu$m
sources. As 22~$\mu$m is closer to the dust emission peak and is not affected by
PAH emission features, it is interesting to compare with the 12$\mu$m galaxies
of our previous analysis. Figure~\ref{fig:ltir_sfr_d4000} shows the linear fits
for the 22~$\mu$m subsample (triangle markers). These relations are very similar
to the 12$\mu$m fits (solid line for SF galaxies, square markers for AGN and
composite galaxies), supporting the independence of our results on the
particularities of a single mid-IR band.

\subsection{Specific Star Formation Rate}
\label{sec:ssfr}
Given the strong correlations between optical or IR light and stellar mass, a
more interesting metric for comparison is the specific star formation rate,
SSFR or SFR/M$_{\odot}$, that measures the current relative to past SFR needed
to build up the stellar mass of the galaxy. The SSFR traces the star formation
efficiency and its inverse defines the timescale for galaxy formation or the
time the galaxy required to build up its current mass. Higher values of SSFR are
indicative of a larger fraction of stars being formed recently. While ideally we
would use gas mass or total mass instead of stellar mass for the normalization,
such masses are not easily measured.

\begin{figure}
\epsscale{1.0}
\plotone{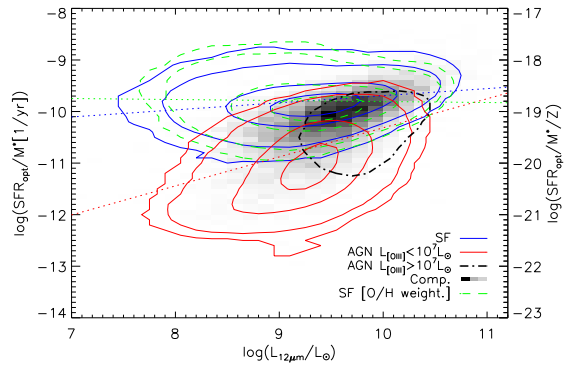}
\caption{Specific star formation rate (SSFR) as a function of 12\,$\mu$m
luminosity for SF galaxies (blue), AGN (red), and composite galaxies
(grayscale). The metallicity-weighted SSFR for star forming galaxies is
indicated by the dashed green contours and its corresponding y-axis is shown on
the right, where Z$=$12+log(OH). To guide the comparison of different
distributions we show simple linear fits of the form
log(SFR)=a$\times$log(L$_{\rm 12\mu m}$)+b (dotted lines). Contours enclose
98\%, 95\%, 68\% and 33\% of the density distribution (68\% for the case of
strong AGN).}
\label{fig:ltir_ssfr_sptype_metal}
\end{figure}

Figure~\ref{fig:ltir_ssfr_sptype_metal} shows the SSFR as a function of 
12\,$\mu$m luminosity for the different galaxy classes. The nearly flat
correlation for SF galaxies means that no matter the IR output, the amount of
star formation per unit mass remains relatively constant. SF galaxies
display a weak dependence with $L_{\rm 12\mu m}$ that gets narrower towards
higher luminosities. As noted before, a possible origin for such residual SSFR
could be due to a metallicity gradient. \citet{calzetti07} studied individual
star forming regions of fixed aperture in
nearby galaxies with known Pa$\alpha$ surface density, and found that low
metallicity galaxies have a small deficit in 24~$\mu$m emission compared to high
metallicity galaxies. \citet{relano} confirmed that while 24~$\mu$m luminosity
is a good metallicity-independent tracer for the SFR of individual HII regions,
the metallicity effect should be taken into account when analyzing SFRs
integrated over the whole galaxy. We test qualitatively for a metallicity effect
by calculating the SFR per unit mass per unit metallicity, where the metallicity
is given by the 12+log(O/H) gas-phase oxygen abundance derived from optical
nebular emission lines. The SF population displays an almost perfectly flat
relationship over almost 4 orders of magnitude in $L_{\rm 12\mu m}$ such that
independent of IR output, a galaxy of given mass and metal content converts
gas into stars at a nearly constant rate. We note that these metallicities
represent the current metal abundance rather than the luminosity-weighted
average of past stellar populations. They also do not suffer from complications
due to $\alpha$-element enhancement or age uncertainties, characteristic of
methods relying on absorption-line indices. \citet{bond} arrive at a similar
conclusion regarding a constant SSFR in nearby galaxies using Herschel
250\,$\mu$m and WISE 3.4\,$\mu$m data.

Compared to SF galaxies, AGN have SSFRs lower by a factor of $\sim$10, mainly
because of their higher stellar masses. However, strong AGN lie much closer to
the SF sequence than weak AGN. The former are hosted by high stellar mass
galaxies, but also have young stellar populations that drive up the SFR at
high $L_{\rm 12\mu m}$. Weak AGN do not have this boost in SFR and hence have
lower SSFRs. Once again, composite systems populate a region intermediate
between SF galaxies and strong AGN. Previous studies (e.g. \citealt{brinchman};
\citealt{salim07}) have shown the relationship between the SFR and $M^*$,
identifying two different sequences: galaxies on a star-forming sequence and
galaxies with little or no detectable SF. While the general result is that the
SSFR of massive, red galaxies is lower at $0<z<3$, the exact dependence of SSFR
on mass is still a matter of debate, particularly in view of recent results that
trace the evolution at higher redshift (e.g. \citealt{noeske}; \citealt{dunne};
but see the discussion by \citealt{rodighiero}). In our sample of SF galaxies we
find that the dependence of $L_{\rm 12\mu m}$ on $M^*$ is such that the
efficiency by which gas is transformed into stars is nearly independent of the
IR emission reprocessed by the dust. In strong AGN the star formation activity
is ``suppressed'' moderately, but considerably more in the case of weak AGN.
This suggests a sequence where a strong-AGN phase is a continuation of the SF
sequence at high stellar mass, that gradually turns AGN into a population with
weakened SF activity and lower $L_{\rm 12\mu m}$, dominated by older and redder
stars.

Based on optical data, \citet{kauff03agn} found that strong AGN hosts are indeed
populated by relatively young stars, suggesting many of them could be
post-starburst systems with extended star formation. With UV data,
\citet{salim07} showed there is a close connection between massive SF galaxies
and strong AGN. Using IR data, we find that the smooth sequence of galaxies from
Figures \ref{fig:ltir_sfr_d4000} and \ref{fig:ltir_ssfr_sptype_metal} supports
the hypothesis of strong AGN being the continuation at the massive-end of the
normal SF sequence. An interesting question is whether the mid-IR luminosity in
powerful AGN is driven by ``normal'' ongoing SF, or by hot dust left over
\emph{after} the last episode of SF. Further investigation is required to fully
understand this matter.

\begin{figure}
\epsscale{1.0}
\plotone{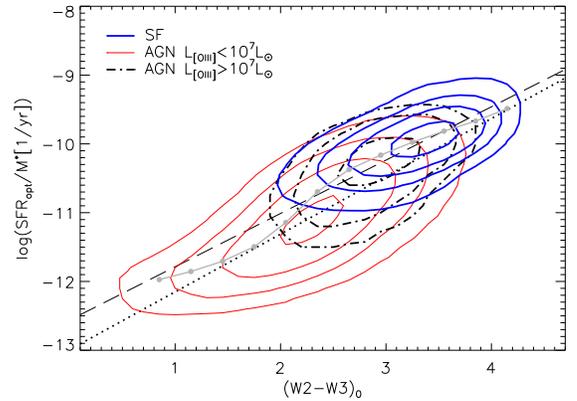}
\caption{Specific star formation rate as a function of restframe 4.6-12\,$\mu$m
galaxy color (W2-W3)$_0$ for SF galaxies (blue), strong AGN (dash-dot black)
and weak AGN (red). The gray line indicates the median relation of the complete
sample. The dashed and dotted lines correspond to linear fits to all sources and
to all AGN, respectively. Contours enclose 95\%, 85\%, 68\% and 33\% of the
density distribution (85\%, 68\% and 33\% for the case of strong AGN).}
\label{fig:ssfr_w2w3_sptype}
\end{figure}

\subsection{SSFR Dependence on Mid-IR Color}
Recent work on resolved nearby galaxies has shown a definite correlation between
IR color and luminosity. \citet{shi} found that for a variety of sources ranging
from ULIRGs to blue compact dwarf galaxies, the flux ratio 
$f_{24\mu m}/f_{5.8\mu m}$ traces the SSFR and also correlates with the
compactness of star forming regions. While ideally we would like to know the
SFR surface density, we first explore the relation between SSFR and 
4.6--12\,$\mu$m galaxy color, as shown in Figure \ref{fig:ssfr_w2w3_sptype}.
Galaxies from the main SF sequence (blue contours) correlate strongly with IR
color, with strong AGN (black contours) continuing the trend at bluer colors.
Weak AGN extend (red contours) that relationship remarkably well toward the low
star formation end, albeit with higher dispersion and a slightly steeper slope.
Hence, for the same increase in SSFR, AGN experience a smaller variation in IR
color than typical SF objects. This is probably due to the combination of a
metallicity effect and the different stellar populations that regulate the IR
emission budget. In any case, this suggests that the higher the SSFR, the more
prominent the 12$\mu$m IR emission becomes relative to 4.6~$\mu$m, where the
latter is expected to strongly correlate with stellar mass. This shows that the
4.6--12\,$\mu$m color serves well as a rough first-order indicator of star
formation activity over three orders of magnitude in SFR. A simple expression
fitting all galaxies is given by
 \begin{equation}
\log SSFR=(0.775\pm0.002)(W2-W3)_{0}-(12.561\pm0.006).
\end{equation}
If the galaxy is known to host an AGN, the more accurate expression is
 \begin{equation}
\log SSFR=(0.840\pm0.008)(W2-W3)_{0}-(12.991\pm0.020).
\end{equation}
These results show that there is a tight link between stellar mass, current star
formation rate and IR color in SF galaxies, which emphasizes the role of the
dominant stellar population in regulating star formation.

\subsection{Effect of AGN on the Energy Budget}
\label{sec:agn_effect}
\begin{figure}
\epsscale{1.0}
\plotone{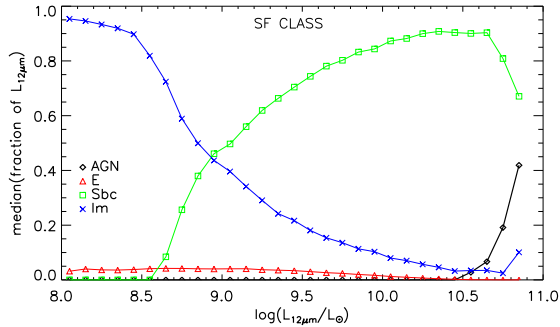}
\caption{Median of the 12\,$\mu$m luminosity fraction contributed by the four
templates from \citet{assef} that are fitted to SDSS $ugriz$ photometry and WISE
3.4~$\mu$m, 4.6~$\mu$m and 12\,$\mu$m fluxes (we also use 22\,$\mu$m data
when available). We include only objects classified as SF from the BPT diagram.}
\label{fig:atype_sf}
\end{figure}

\begin{figure}
\epsscale{1.0}
\plotone{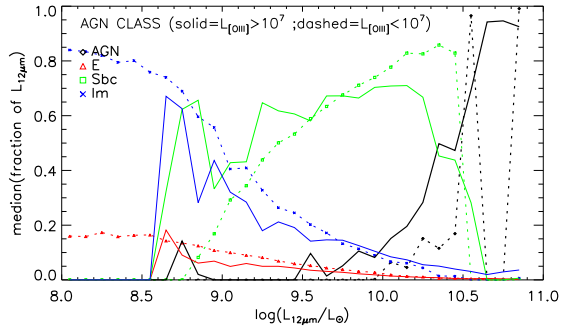}
\caption{Same as Figure~\ref{fig:atype_sf}, but for objects classified as weak
AGN (dashed) and strong AGN (solid).}
\label{fig:atype_agn}
\end{figure}

In the previous sections we noted that the emission from the AGN could have a
significant effect on the IR emission, and potentially bias the luminosities of
the AGN galaxy class. This could be particularly important for low SFR galaxies.
We test for this effect by estimating the contribution of the AGN to the total
energy budget for sources classified from the BPT diagram as either an AGN or as
an SF galaxy. To do so, we analyze the fraction of the 12$\mu$m luminosity
contributed by each of the four templates used in the SED fitting process to our
optical+IR photometry, paying particular attention to the AGN component.
Figure~\ref{fig:atype_sf} shows the median fraction of 12\,$\mu$m luminosity
contributed by each template, for objects classified as SF galaxies. The
majority of the power is split among the irregular (Im), spiral (Sbc) and
elliptical (E) templates, though the AGN contribution becomes significant
for the most luminous sources, above $10^{10.8}L_{\odot}$. This implies that
the AGN has a negligible influence in SF galaxies. Figure~\ref{fig:atype_agn}
shows these fractions for weak and strong AGN galaxy classes. The elliptical
component is now more prominent in weak AGN of low luminosity, which is not
unexpected. In general, the AGN component is now more important but is still far
from contributing significantly below $10^{10.8}L_{\odot}$. In most weak AGN
($\sim$80\%) the AGN contribution to the 12\,$\mu$m luminosity is below 40\%.
About $\sim$70\% of strong AGN show similar low AGN contributions at
12\,$\mu$m. We note that in Figure~\ref{fig:atype_agn} we used WISE aperture
photometry for extended, nearby sources and profile-fitting magnitudes for
unresolved galaxies with $\chi^2<3$ (see Section 4.5 of WISE Preliminary Release
Explanatory Supplement for further details). Although the differences are small,
aperture photometry improves the quality of the SED fit of low luminosity
galaxies, as it captures the more extended flux of objects at low redshifts.

\begin{figure*}
\epsscale{1.0}
\plottwo{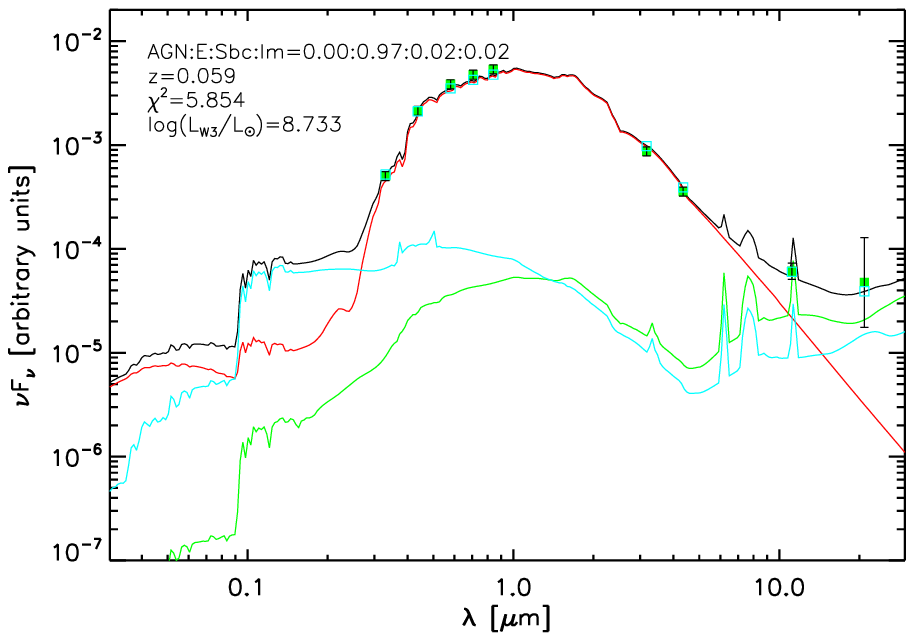}{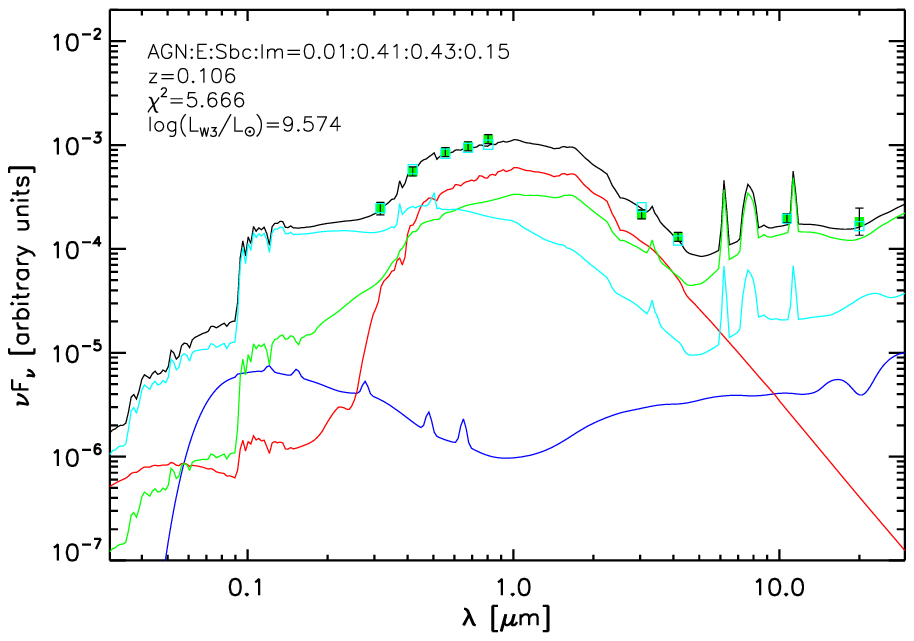}
\caption{SED fits for AGN with the median $\chi^2$ in two bins of
$L_{\rm[OIII]}$ centered at $L_{\rm[OIII]}\sim10^6\, L_\odot$ (left) and
$L_{\rm[OIII]}\sim10^7\, L_\odot$ (right). The labels indicates the fraction of
the bolometric luminosity contributed by each component template, which are
shown in blue (AGN), red (E), green (Sbc), cyan (Im). The black line shows the
total SED. The model SDSS and WISE photometry (cyan squares) are very similar to
the observed photometric points (green squares).}
\label{fig:seds_agn}
\end{figure*}

Note that in both cases, SF galaxies and AGN, the emission is dominated by the
spiral (Sbc) component, which has a relatively high mid-IR SED. This is because
the template, originally constructed from \citet{coleman} and extended into the
UV and IR with \citet{bruzual03} synthesis models, also considers emission in
the mid-IR from dust and polycyclic aromatic hydrocarbons (PAHs). These are
added by ad hoc linear combinations of appropriate parts of NGC 4429 and M82
SEDs obtained by \citet{devriendt}. Figure~\ref{fig:seds_agn} shows the SED fit
for AGN with luminosities of $L_{\rm[OIII]}\sim10^6\, L_\odot$ and
$L_{\rm[OIII]}\sim10^7\, L_\odot$. In each case we plot the object with the
median $\chi^2$, i.e. the typical fit for sources of those luminosities. The 9
photometric bands are well fitted by the model in most cases. For AGN with
$L_{\rm[OIII]}>10^{7.5}\, L_\odot$ we find the SED fit is reasonably good
(although in general with higher $\chi^2$) except
at the 22\,$\mu$m band. We believe this is caused by the limitation of the
algorithm (not the templates themselves) to properly fit highly reddened AGN
fainter than their hosts, as it is designed to punish the excessive use of
reddening on the AGN when few relevant data points are used. Modifying the
algorithm slightly to remove this behavior, we are able to obtain fits with
better $\chi^2$ values for these objects assigning much higher AGN fractions and
reddening values, yet the lack of farther IR data to determine the origin of the
22\,$\mu$m excess makes these numbers also uncertain. Therefore, while these
results do not definitely rule out that the central AGN could have a
considerable effect in some extreme sources (e.g. the very strong AGN), it
certainly shows that they are not relevant for most of the galaxy populations
analyzed in this paper.


\section{Summary}
In this work we have taken advantage of recently released data from WISE and
SDSS to construct the largest IR-optical sample of galaxies with 12\,$\mu$m
fluxes and optical spectra available at $\langle z \rangle \sim 0.1$. This
sample allowed us to investigate with high statistical significance how physical
parameters such as color, stellar mass, metallicity, redshift, and SFR of
12\,$\mu$m-selected galaxies compare with purely optically selected samples. We
have quantified how the SFR estimates compare for the different spectral types
as a function of stellar mass, galaxy age and IR color in order to pinpoint the
underlying source of 12\,$\mu$m emission and therefore up to what extent it
could be interpreted as a useful SFR indicator.

The main results of this paper can be summarized as follows:
\begin{itemize}
\item
In general, the WISE-SDSS 12\,$\mu$m-selected galaxy population traces the blue,
late-type, low mass sequence of the bimodal galaxy distribution in the local
Universe. It also traces intermediate-type objects with active nuclei, avoiding
the bulk of the red and ``dead'' galaxies without emission lines. Most sources
have normal to LIRG luminosities, but (few) ULIRGs are also present.
\item
The IR emission of SF galaxies and strong AGN, dominated by the blue, young
stellar population component, is well correlated with the optical SFR. There is
a small tendency of low SFR systems to have slighly lower IR luminosity when
compared to the canonical relation of \citet{charyelbaz}. These are low SFR, low
mass systems that likely become more transparent due to the increasing
fraction of light that escapes unabsorbed and hence suppresses $L_{\rm 12\mu
m}$. However other effects like the dust distribution or metallicity could be
relevant as well. The latter is shown by the (weak) SSFR dependence on $L_{\rm
12\mu m}$. In general, the mid-IR emission at 22\,$\mu$m follows similar
correlations seen for the 12\,$\mu$m-selected sample, suggesting that these
results do not critically depend on a single IR band.
\item
SF galaxies are forming stars at an approximately constant rate per unit mass
for an IR output ranging over five orders of magnitude. There is small tendency
for more luminous objects to have enhanced SSFR, which could be interpreted as a
sign of SF histories peaking toward later times. However, this residual
dependence seems to be caused by a metallicity gradient. Once factored out, the
relationship becomes nearly flat. Strong AGN behave as a continuation at the
massive-end of the normal SF sequence, where the AGN (possibly after a starburst
episode) gradually quenches SF and weakens as it consumes the gas available,
with the mid-IR emission fading in consequence.
\item
The mid-IR 4.6--12\,$\mu$m restframe color can be used as a first-order
indicator of the overall SF activity in a galaxy, as it correlates well with the
specific SFR. For increasing SFR/M$^*$, the IR emission becomes more prominent
at 12\,$\mu$m (associated with dust emission) relative to 4.6~$\mu$m (associated
with stellar mass). 
\item
For the case of SF galaxies, most of the mid-IR luminosity distribution is
concentrated in systems younger than $\sim$0.5~Gyr. Redder galaxies are
dominated by older stellar populations, which contribute increasingly to the
12\,$\mu$m emission. While many of these galaxies host an AGN (usually weak) the
12\,$\mu$m energy budget is generally not dominated by the central active
nuclei. This might well not be the case of bright galaxies with very strong
active nuclei ($L_{\rm[OIII]}>10^7.5\, L_\odot$) where a considerably larger
fraction of mid-IR emission could be due to the AGN. Spatially resolved, longer
wavelength IR data and further modeling is necessary to fully understand these
sources.
\end{itemize}

\acknowledgments
The authors thank G. Kauffmann, J. Brinchmann and S. Salim for useful
suggestions. This publication makes use of data products from the Wide-field
Infrared Survey Explorer, which is a joint project of the University of
California, Los Angeles, and the Jet Propulsion Laboratory/California Institute
of Technology, funded by the National Aeronautics and Space Administration.
Funding for the SDSS and SDSS-II has been provided by the Alfred P. Sloan
Foundation, the Participating Institutions, the National Science Foundation, the
US Department of Energy, the National Aeronautics and Space Administration, the
Japanese Monbukagakusho, the Max Planck Society and the Higher Education Funding
Council for England. The SDSS Web Site is http://www.sdss.org/. R.J.A. was
supported by an appointment to the NASA Postdoctoral Program at the Jet
Propulsion Laboratory, administered by Oak Ridge Associated Universities through
a contract with NASA.


\begin{thebibliography}{99}
\bibitem[\protect\citeauthoryear{Abazajian et~al.}{2009}]{abazajian} Abazajian 
K. N. et al., 2009, ApJS, 182, 543
\bibitem[\protect\citeauthoryear{Alonso-Herrero et~al.}{1997}]{alonsoh97}
Alonso-Herrero A.; Ward M. J., Kotilainen J. K., 1997, MNRAS, 288, 977
\bibitem[\protect\citeauthoryear{Alonso-Herrero et~al.}{2006}]{alonsoh}
Alonso-Herrero A., Rieke G. H., Rieke M. J., Colina L., Perez-Gonzalez P. G.,
Ryder S. D., 2006, ApJ, 650, 835
\bibitem[\protect\citeauthoryear{Assef et~al.}{2010}]{assef} Assef R. J. et
al., 2010, ApJ, 713, 970
\bibitem[\protect\citeauthoryear{Baldwin, Phillips \& Terlevich}{1981}]{bpt}
Baldwin J., Phillips M., Terlevich R., 1981, PASP, 93, 5
\bibitem[\protect\citeauthoryear{Baldry et~al.}{2004}]{baldry} Baldry I. K.,
Glazebrook K., Brinkmann J., Ivezic Z., Lupton R. H., Nichol R. C.,
Szalay A. S., 2004, ApJ, 600, 681
\bibitem[\protect\citeauthoryear{Blanton \& Roweis}{2007}]{blanton} Blanton 
M. R., Roweis S., 2007, AJ, 133, 734
\bibitem[\protect\citeauthoryear{Bond et~al.}{2011}]{bond} Bond N. A. et al.,
2011, in preparation
\bibitem[\protect\citeauthoryear{Brinchmann et~al.}{2004}]{brinchman}
Brinchmann J., Charlot S., White S. D. M., Tremonti C., Kauffmann G., Heckman
T., Brinkmann J., 2004, MNRAS, 351, 1151
\bibitem[\protect\citeauthoryear{Bruzual \& Charlot}{1993}]{bruzual93} Bruzual
G., Charlot S., 1993, ApJ, 405, 538
\bibitem[\protect\citeauthoryear{Bruzual \& Charlot}{2003}]{bruzual03} Bruzual
G., Charlot S., 2003, MNRAS, 344, 1000
\bibitem[\protect\citeauthoryear{Calzetti et~al.}{2007}]{calzetti07} Calzetti
D. et al. 2007, ApJ, 666, 870
\bibitem[\protect\citeauthoryear{Charlot \& Fall}{2000}]{charlot_fall} Charlot
S., Fall S. M., 2000, ApJ, 539, 718
\bibitem[\protect\citeauthoryear{Charlot \& Bruzual}{2008}]{charlot08}
Charlot S., Bruzual G., 2008, in preparation
\bibitem[\protect\citeauthoryear{Chary \& Elbaz}{2001}]{charyelbaz} Chary R.,
Elbaz D., 2001, ApJ, 556, 562
\bibitem[\protect\citeauthoryear{Coleman et~al.}{1980}]{coleman} Coleman G. D.,
Wu C.-C., Weedman D. W., 1980, ApJS, 43, 393
\bibitem[\protect\citeauthoryear{Cutri et~al.}{2011}]{cutri} Cutri R. et al.,
2011, WISE Explanatory Supplement.
\bibitem[\protect\citeauthoryear{Daddi et~al.}{2007}]{daddi} Daddi E. et al.,
2007, ApJ, 670, 156
\bibitem[\protect\citeauthoryear{Devriendt et~al.}{1999}]{devriendt} Devriendt
J. E. G., Guiderdoni B., Sadat R., 1999, A\&A, 350, 381
\bibitem[\protect\citeauthoryear{Duc et~al.}{2002}]{duc} Duc P.-A. et al., 2002,
A\%A, 382, 60
\bibitem[\protect\citeauthoryear{Dunne et~al.}{2009}]{dunne} Dunne L. et al.,
2009, MNRAS, 394, 3
\bibitem[\protect\citeauthoryear{Eisenhardt et~al.}{2011}]{eisenhardt}
Eisenhardt P. et al., 2011, in preparation
\bibitem[\protect\citeauthoryear{Ferland}{1996}]{ferland} Ferland G., 1996,
Ferland G., 1996, Hazy: A Brief Introduction to CLOUDY. Internal Report, Univ.
Kentucky
\bibitem[\protect\citeauthoryear{Griffith et~al.}{2011}]{griffith} Griffith R.
L. et al.,2011, ApJ, 736, 22
\bibitem[\protect\citeauthoryear{Heckman et~al.}{1998}]{heckman} Heckman T. M.,
Robert C., Leitherer C., Garnett D. R., van der Rydt F., 1998, ApJ, 503, 646
\bibitem[\protect\citeauthoryear{Heckman et~al.}{2004}]{heckman04} Heckman, T.
M., Kauffmann G., Brinchmann J., Charlot S., Tremonti C., White, S. D. M., 2004,
ApJ, 613, 109
\bibitem[\protect\citeauthoryear{Hou, Wu \& Han}{2009}]{hou} Hou L. G., Wu
Xue-Bing, Han J. L., 2009, ApJ, 704, 794
\bibitem[\protect\citeauthoryear{Jannuzi et~al.}{2010}]{jannuzi} Jannuzi B. et
al., 2010, Bulletin of the American Astronomical Society Meeting 215, Vol 42.,
p.513
\bibitem[\protect\citeauthoryear{Jarrett et~al.}{2011}]{jarrett} Jarrett T. et
al., 2011, ApJ, 735, 112
\bibitem[\protect\citeauthoryear{Kauffmann et~al.}{2003a}]{kauff03} Kauffmann 
G. et al., 2003a, MNRAS, 341, 33
\bibitem[\protect\citeauthoryear{Kauffmann et~al.}{2003b}]{kauff03agn}
Kauffmann G. et al., 2003b, MNRAS, 346, 1055
\bibitem[\protect\citeauthoryear{Keel et~al.}{1994}]{keel} Keel W. C., De
Grijp M. H. K., Miley G. K., Zheng W., 1994, A\&A, 283, 791
\bibitem[\protect\citeauthoryear{Kelson \& Holden}{2010}]{kelson}
Kelson D. D., Holden B. P., 2010, ApJL, 713, 28
\bibitem[\protect\citeauthoryear{Kennicutt}{1998}]{kennicutt98} Kennicutt, R.
C. Jr., 1998, ARA\&A, 36, 189
\bibitem[\protect\citeauthoryear{Kennicutt et~al.}{2009}]{kennicutt09}
Kennicutt, R. C. Jr. et al. 2009, ApJ, 703, 1672
\bibitem[\protect\citeauthoryear{Kochanek et~al.}{2011}]{kochanek} Kochanek C.
et al., 2011, in preparation
\bibitem[\protect\citeauthoryear{Lake et~al.}{2011}]{lake} Lake S. E. et al.,
2011, in preparation
\bibitem[\protect\citeauthoryear{Leisawitz \& Hauser}{1988}]{leisawitz}
Leisawitz D, Hauser M. G., 1988, ApJ, 332, 954
\bibitem[\protect\citeauthoryear{Martin et~al.}{2007}]{martin} Martin D. C.
et~al., 2007, ApJS, 173, 342
\bibitem[\protect\citeauthoryear{Mulchaey et~al.}{1994}]{mulchaey} Mulchaey J.
S., Koratkar A., Ward M. J., Wilson A. J., Whittle M., Antonucci, R. R. J.,
Kinney A. L., Hurt T., 1994, ApJ, 436, 58
\bibitem[\protect\citeauthoryear{Neugebauer et~al.}{1984}]{neugebauer}
Neugebauer G. et al., 1984, ApJ, 278, 1
\bibitem[\protect\citeauthoryear{Noeske et~al.}{2007}]{noeske} Noeske K. G. 
et~al., 2007, ApJ, 660, L43
\bibitem[\protect\citeauthoryear{Rieke et~al.}{2009}]{rieke} Rieke G.
H., Alonso-Herrero A., Weiner B. J., P\'{e}rez-Gonz\'{a}lez P. G., Blaylock M.,
Donley J. L., Marcillac D., 2009, ApJ, 692, 556
\bibitem[\protect\citeauthoryear{Rela\~{n}o et~al.}{2007}]{relano} Rela\~{n}o
M., Lisenfeld U., P\'{e}rez-Gonz\'{a}lez P. G., V\'{i}lchez J. M., Battaner, E.,
2007, ApJ, 667, 141
\bibitem[\protect\citeauthoryear{Rocca-Volmerange et~al.}{2007}]{roccav} 
Rocca-Volmerange B., de Lapparent V., Seymour N., Fioc M, 2007, A\&A, 475, 801
\bibitem[\protect\citeauthoryear{Rodighiero et~al.}{2010}]{rodighiero}
Rodighiero G. et al., 2010, A\&A, 518, 25
\bibitem[\protect\citeauthoryear{Saintonge et~al.}{2011}]{saintonge} Saintonge
A. et al., 2011, MNRAS, 415, 61
\bibitem[\protect\citeauthoryear{Salim et~al.}{2007}]{salim07} Salim S. et al.,
2007, ApJS, 173, 267
\bibitem[\protect\citeauthoryear{Salim et~al.}{2009}]{salim09} Salim S. et al.,
2009, AJ, 700, 161
\bibitem[\protect\citeauthoryear{Schlegel, Finkbeiner \& Davis}{1998}]{schlegel}
Schlegel D. J., Finkbeiner D. P., Davis M., 1998, AJ, 500, 525
\bibitem[\protect\citeauthoryear{Seymour et~al.}{2007}]{seymour} Seymour N.,
Rocca-Volmerange B., de Lapparent V., 2007, A\&A, 475, 791
\bibitem[\protect\citeauthoryear{Shi et~al.}{2011}]{shi} Shi Y. et al., 2011,
in preparation
\bibitem[\protect\citeauthoryear{Shimasaku et~al.}{2001}]{shimasaku} Shimasaku
K. et al., 2001, AJ, 122, 1238
\bibitem[\protect\citeauthoryear{Skrutskie et~al.}{2006}]{skrutskie} Skrutskie
M.F. et al., 2006, AJ, 131, 1163
\bibitem[\protect\citeauthoryear{Soifer et~al.}{1987}]{soifer} Soifer B. T. et
al., 1987, ApJ, 320, 238
\bibitem[\protect\citeauthoryear{Spinoglio \& Malkan}{1989}]{spinoglio}
Spinoglio L., Malkan M. A., 1989, AJ, 342, 83
\bibitem[\protect\citeauthoryear{Stern et~al.}{2011}]{stern2011} Stern D. et
al., 2011, in preparation
\bibitem[\protect\citeauthoryear{Stoughton et~al.}{2002}]{stoughton} Stoughton
C. et al., 2002, AJ, 123, 485
\bibitem[\protect\citeauthoryear{Strateva et~al.}{2001}]{strateva} Strateva I.
et al., 2001, AJ, 122, 1861
\bibitem[\protect\citeauthoryear{Strauss et~al.}{2002}]{strauss} Strauss M. A.
et al., 2002, AJ, 124, 1810
\bibitem[\protect\citeauthoryear{Teplitz et~al.}{2011}]{teplitz} Teplitz H. I.
et al., 2011, AJ, 141, 1
\bibitem[\protect\citeauthoryear{Tremonti et~al.}{2004}]{tremonti} Tremonti C.
A. et~al., 2004, AJ, 613, 898
\bibitem[\protect\citeauthoryear{Wright et~al.}{2010}]{wright} Wright E.~L.
et~al., 2010, AJ, 140, 1868
\bibitem[\protect\citeauthoryear{Wu et~al.}{2005}]{wu} Wu H. et~al., 2005,
ApJ, 632, 79
\bibitem[\protect\citeauthoryear{Yan et~al.}{2011}]{yan} Yan L. et al., 2011,
in preparation
\bibitem[\protect\citeauthoryear{York et~al.}{2000}]{york} York D.~G. et al.,
2000, AJ, 120, 1579
\bibitem[\protect\citeauthoryear{Zhu et~al.}{2008}]{zhu} Zhu Y. N., Wu H., Cao
C., Li H.N., 2008, ApJ, 686, 155
\end{thebibliography}
\end{document}